\documentclass[journal,letterpaper]{IEEEtran_nssmic}

\usepackage{cite}      

\usepackage{graphicx}  
\begin{document}
%
\title{Evaluation of 5 mm-thick CdTe Detectors from the Company Acrorad}
%
%
\author{Alfred~Garson~III$^{1}$,        
        Ira~V.~Jung$^{1}$, Jeremy~Perkins$^{1}$, and~Henric~Krawczynski$^{1}$%
\thanks{$^{1}$Washington University in St. Louis, 1 Brookings Drive, CB 
    1105, St.\ Louis, Mo, 63130}
\thanks{A. Garson:agarson3@hbar.wustl.edu}
\thanks{Manuscript received November 11, 2005.}}

\maketitle

\begin{abstract}
  Using 2$\times$2$\times$0.5 cm$^3$ Cadmium Telluride (CdTe) substrates 
  from the company Acrorad, we have fabricated detectors with planar cathode contacts 
  and 8$\times$8 anode pixels. We investigate the I-V characteristics and energy resolution
  of the detectors for different contact materials and surface treatments. 
  After biasing the detectors for a certain time, the dark currents increase dramatically.
  Our studies show that the time before breakdown decreases for higher detector temperatures
  and for higher applied bias voltages. We obtained the best results with a Pt cathode 
  contact and In anode pixels when we heat the detector to 90$^\circ$C for 30 minutes prior
  to depositing the In contacts. Flood-illuminating the detector with 662 keV X-rays, 
  we measured the pulse length distribution and derived an electron mobility of $\sim$~860 cm$^{2}$ V$^{-1}$ s$^{-1}$.
  We show that the energy resolution can be improved by correcting the anode signals
  for the depth of the primary interaction. Operated at a temperature of -40$^\circ$C 
  and a cathode bias voltage of -500 V, the best pixels of the best detector achieved 
  full width half maximum (FWHM) energy resolutions of 8 keV (6.4\%) and
  23 keV (3.4\%) at 122 keV and  662 keV, respectively. 
\end{abstract}


\section{Introduction}
%
%
%
%
\PARstart{C}{admium} Telluride (CdTe) and Cadmium Zinc Telluride (CZT) detectors 
are having a major impact on the field of hard X-ray astronomy. 
The Swift\cite{Gehr} satellite carries the Burst Alert Telescope (BAT) 
that uses CZT detectors. The International Gamma-Ray Astrophysics 
Laboratory (INTEGRAL)\cite{Uber} includes the Imager on Board the 
Integral Satellite (IBIS) which is built with CdTe detectors. 
NASA's Beyond Einstein roadmap included two X-ray missions, 
Constellation-X\cite{ConX} and the Black Hole Finder Probe (BHFP). 
CZT detectors will play an important role for both missions. 
The Energetic X-ray Imaging Survey Telescope (EXIST)\cite{Grin} is a strong candidate 
for the BHFP and uses pixelated Si and 0.5 cm thick CZT detectors and the 
coded mask approach to cover the energy range from 5 keV to 600 keV. 
The Non-thermal Energy eXploration Telescope (NeXT)\cite{NeX1}\cite{NeX2} 
mission proposed in Japan is a successor to the Suzaku mission and will 
utilize CdTe on two of its instruments. 
The Hard X-ray Imager (HXI) on board NeXT will cover the energy range 
from 8 to 80 keV using a Si CCD combined with a pixelated CdTe detector. 
NeXT will also carry the Soft-Gamma Detector (SGD), which uses CdTe for 
an extended energy range up to 1 MeV. The broad energy ranges of 
these missions requires either thick detectors or layers of thin detectors. 
In this presentation, we focus on 5 mm-thick CdTe detectors.

While CdTe detectors show at room temperature a poorer 
performance than CZT detectors, CdTe might be an alternative to CZT in spaceborne applications where temperatures 
around -20$^\circ$C can be achieved without the need for cryogenic cooling. 
Several groups tested relatively thin CdTe detectors (e.g. \cite{Taka,Oonuki,Gnat,Aoki}).
We report here on results from evaluating 0.5 cm thick CdTe detectors.
Using substrates from the company Acrorad \cite{Acrorad}, we deposited
different cathode and anode contacts and measured the I-V characteristics,
the energy resolutions, and the electron mobility. 
%
\section{Detector Fabrication}
We purchased two uncontacted 2$\times$2$\times$0.5 cm$^3$ CdTe substrates from Acrorad. 
In the following, we refer to the two detectors as D1 and D2. 
The substrates were cleaned in DI water and methanol. Prior to contact deposition, 
the detectors were polished with three grades of abrasive paper followed by 0.3 and 
0.05 $\mu$m alumina suspension. Optionally, the detectors were etched for 2 minutes 
in a solution of 5\% Br in 95\% methanol. We deposited planar cathode contacts 
and $8 \times 8$ anode pixels on the CdTe substrates with an electron beam evaporator. 
The 1.6 mm diameter pixels with 2.4 mm pixel pitch were deposited using a 
0.3~mm brass foil shadow mask. The first detectors showed modest performances. 
We thus started to use a method similar to the one described by Takahashi et al.\ (1999), 
and heated D1 at 90$^\circ$C (30 minutes) and D2 at 140$^\circ$C (15 minutes) prior to the deposition of the anode contacts.
The heating was performed in a 10$^{-7}$ Torr vacuum with the anode side 
facing an Hg lamp, and a temperature sensor being attached to the 
backside of the detector. As discussed further below, the heat treatment 
 markedly improved the energy resolution achieved with the detectors. 

We used the high-workfunction metals Au and Pt as cathode materials and the 
low-workfunction metal In as anode material. The high-workfunction Au and Pt 
metals are expected to form ohmic contacts on the p-type CdTe and the 
low-workfunction contacts blocking contacts. The heat treatment is thought to 
remove the Te-rich surface layer and to improve the contact properties of the
CdTe--In junctions \cite{Taka}. We will refer to the two detectors as D1a-D1b 
and D2a-D2c, depending on which surface treatments and contact materials were used (see Table 1). 
\section{Results}
\subsection{I-V Characteristics}
We measured the pixel-cathode dark current as a function of the cathode 
bias voltage for detector temperatures between $-40^\circ$C and 20$^\circ$C. Holding all pixels 
at ground potential, the currents were measured for three central pixels.

After keeping the detectors biased at a certain voltage for some time,
we observed the well known breakdown phenomenon: a dramatic increase 
of the dark current. We found that if the bias voltage was set to zero 
after a breakdown and then immediately returned to its previous value, 
the breakdown would occur after a shorter period of time. 
However, if the bias voltage remained at zero for several minutes and 
was then returned to its previous value, the breakdown time would be 
very similar to its original value. 

Figure 1 shows the current for the detector D1a at different temperatures and 
different bias voltages. It is evident that the time before breakdown 
decreases with increasing detector temperature and with more negative 
cathode bias voltages. Figure 2 shows a very similar breakdown 
behaviour for the same crystal, this time depositing the 
In pixels after polishing, etching, and heat treatment (detector D1b).
As discussed further below, our heat treatment may have used
temperatures too low to strongly suppress the breakdown behaviour.
\begin{figure}
\centering
\includegraphics[width=3.5in]{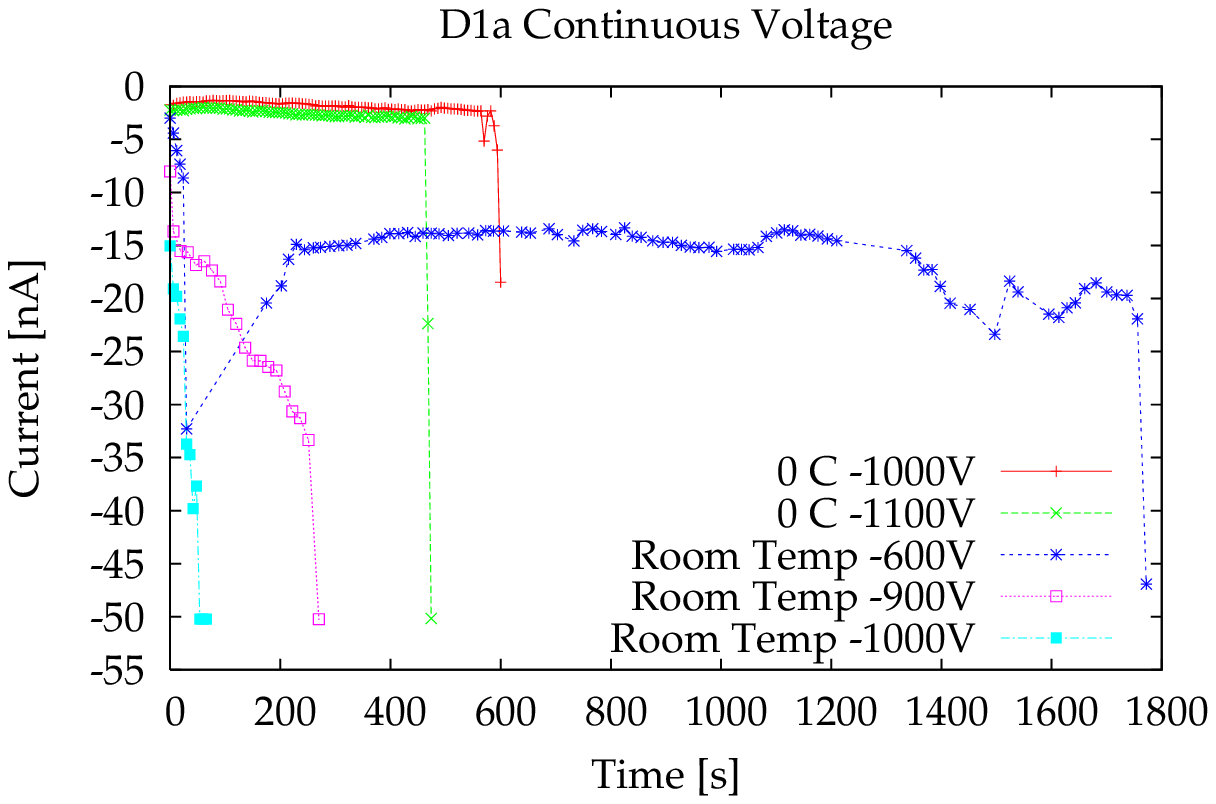}
\caption{Breakdown behaviour for D1a (anode deposition after polishing). 
From left to right, the measurements were taken at 20$^\circ$C -1000 V, 20$^\circ$C -900 V, 
0$^\circ$C -1100 V, 0$^\circ$C -1000 V, and 20$^\circ$C -600 V. }
\label{bkdw}
\end{figure}
\begin{figure}
\centering
\includegraphics[width=3.5in]{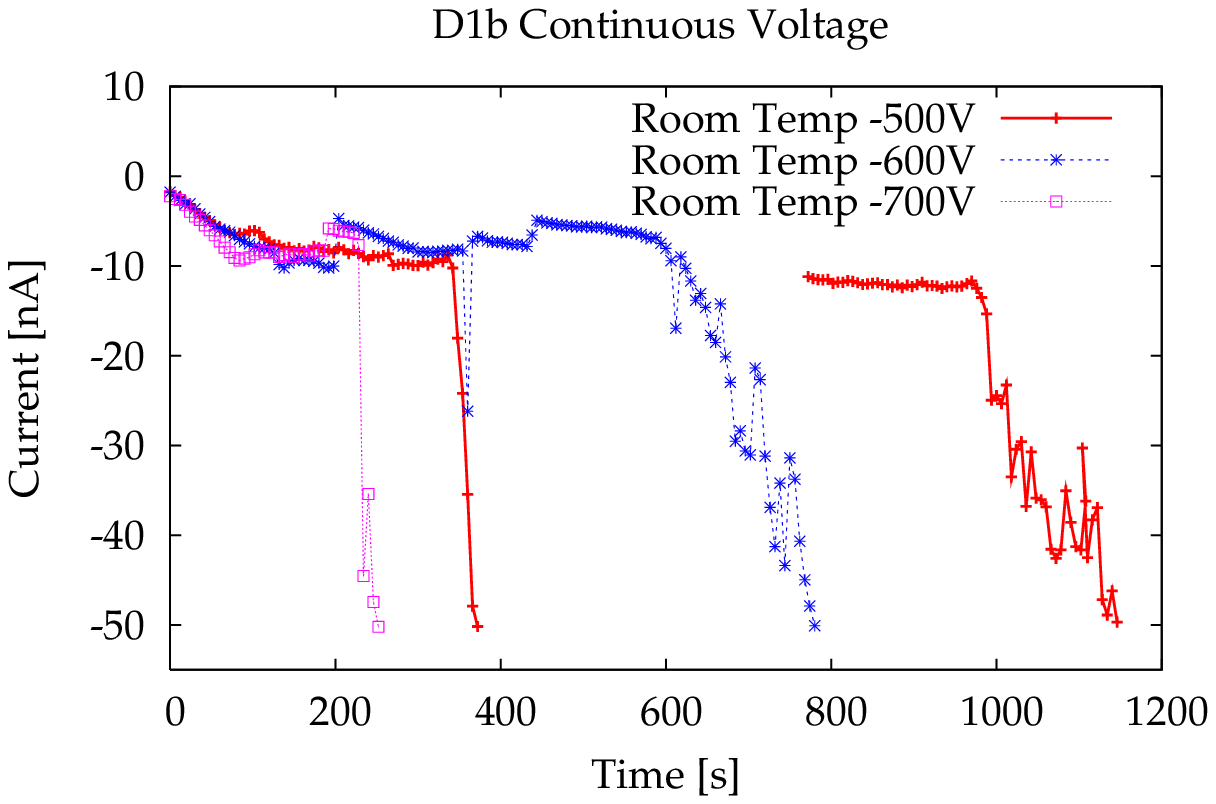}
\caption{Breakdown behaviour for detector D1b (anode deposition after polishing, 2 minutes etching 
and heat treatment at 90$^\circ$C for 30 min). In the case of the -500 V bias, the voltage was
automatically shut down when the current reached -50 nA. After some waiting time, it was set 
to -500 V again. All measurements were taken at room temperature. The anodes were kept at 
ground potential, and the cathode bias voltages were from left to right -700 V, -600 V, and -500 V.}
\label{bkdw2}
\end{figure}

The I-V measurements showed a wide range of variation from detector to detector and from pixel to pixel. 
Figure 3 shows the dark current for the detector D1a as a function of bias voltage for various
temperatures. All measurements were taken before breakdown occurred. 
It can be recognized that the I-V characteristics are non-linear. 
Figure 4 shows the measurements for three pixels on D1b; a large variation from pixel to pixel
can be recognized. Figure 5 compares the results for D2a (Pt cathode) and D2b (Au cathode).
The detector with Au cathode shows very high dark currents that compromise the detector's
spectroscopic performance.
\begin{figure}
\centering
\includegraphics[width=3.5in]{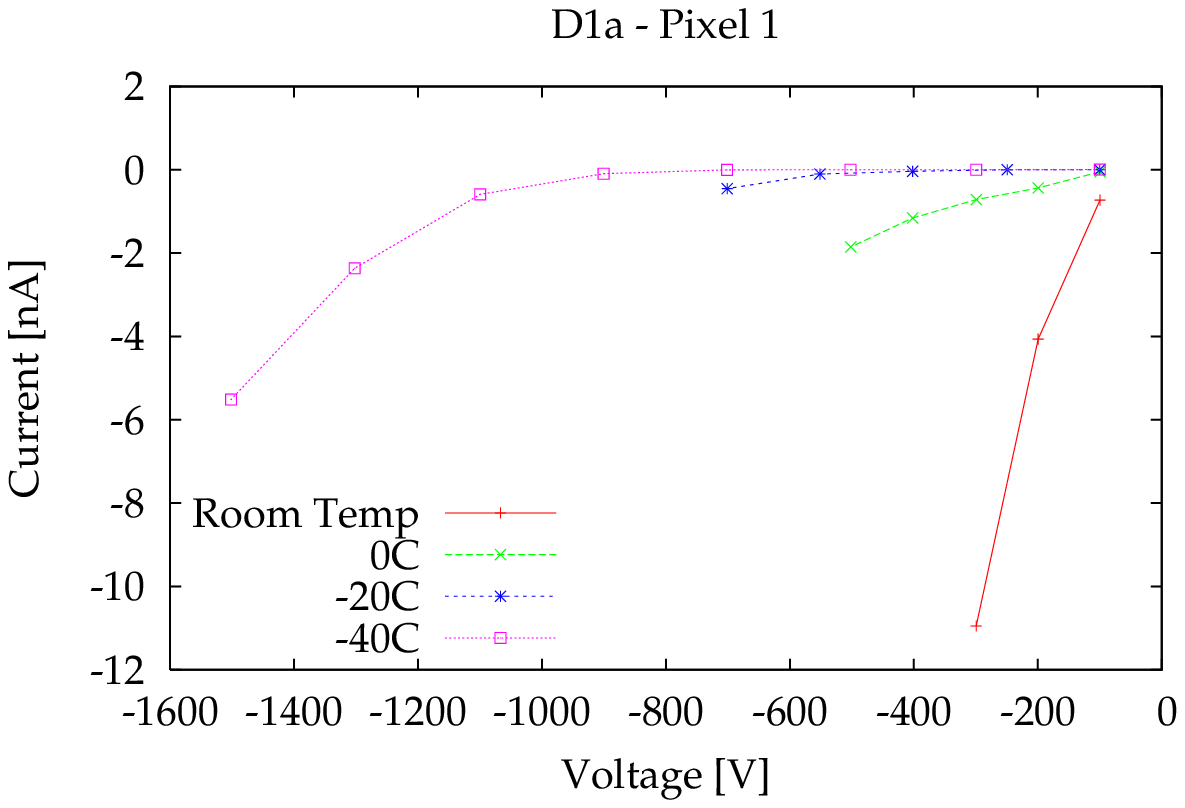}
\caption{
The I-V behaviour at various temperatures for D1a (anode deposition after polishing). 
From top to bottom, measurements were taken at -40$^\circ$C, -20$^\circ$C, 0$^\circ$C, and 20$^\circ$C.}
\label{IV1}

\end{figure}
\begin{figure}
\centering
\includegraphics[width=3.5in]{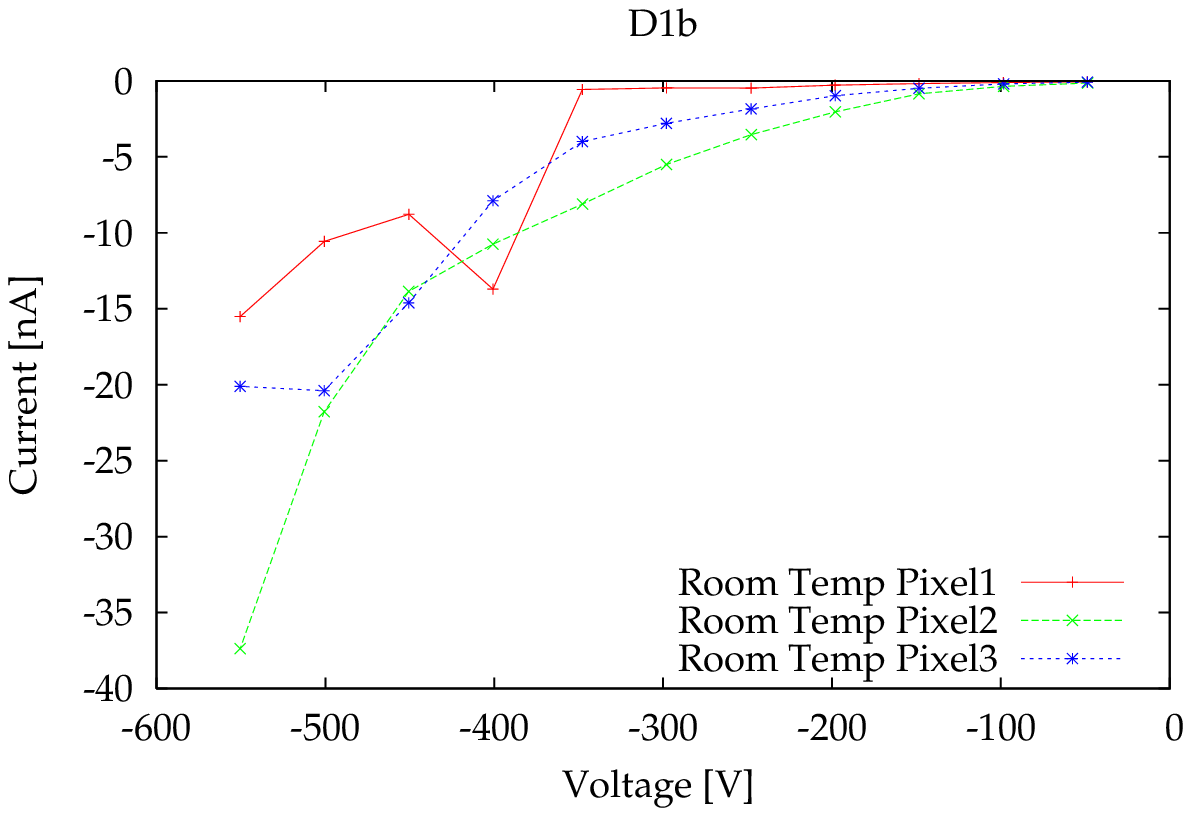}
\caption{
Room temperature I-V results for three pixels of detector D1b (anode deposition after polishing, 2 min etching 
and heat treatment at 90$^\circ$C for 30 min). Different pixels show different behaviours.}
\label{IV2}

\end{figure}
\begin{figure}
\centering
\includegraphics[width=3.5in]{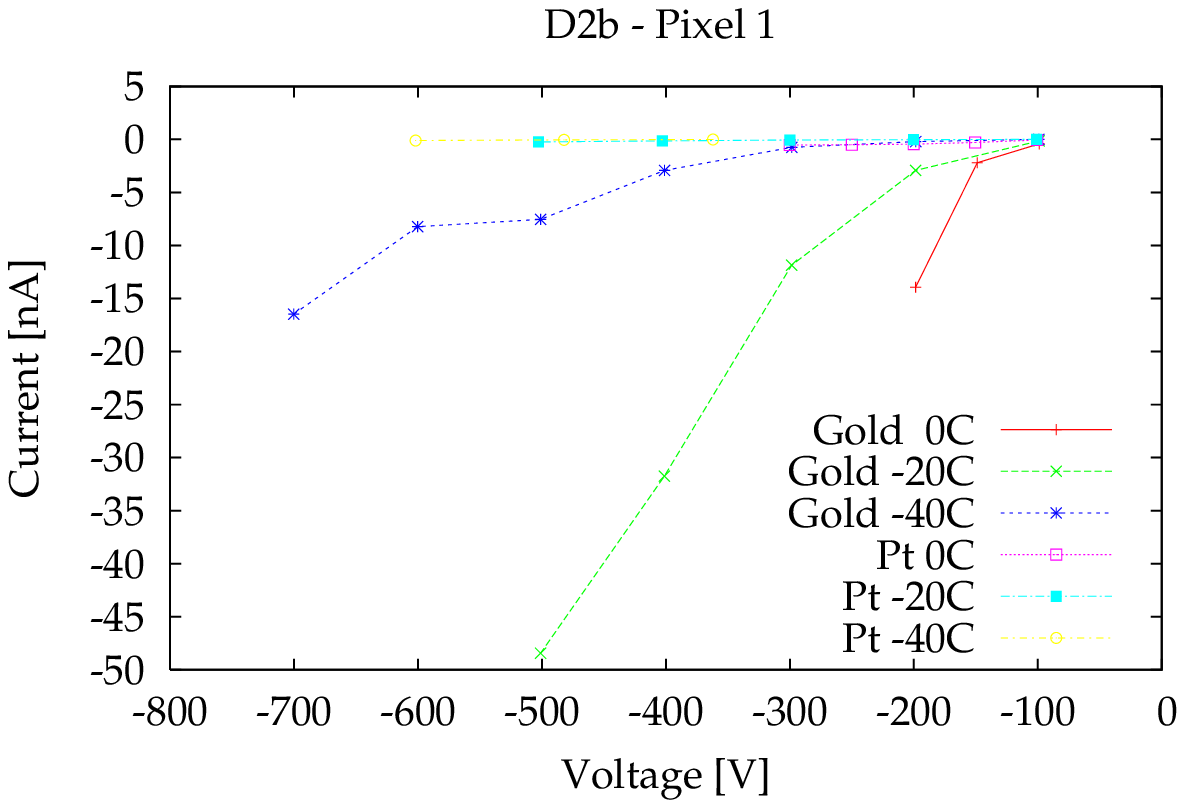}
\caption{
Comparison of I-V behaviour for detector D2 with Pt cathode (D2a) and Au cathode (D2b). 
From top to bottom, measurements were taken at $-40^\circ$C (D2a), -20$^\circ$C (D2a), 0$^\circ$C (D2a), -40$^\circ$C (D2b), -20$^\circ$C (D2b), 0$^\circ$C (D2b), }
\label{D2IV}

\end{figure}
\subsection{Energy Resolution and Mobility Measurements}
\begin{figure}
\centering
\includegraphics[width=2.5in]{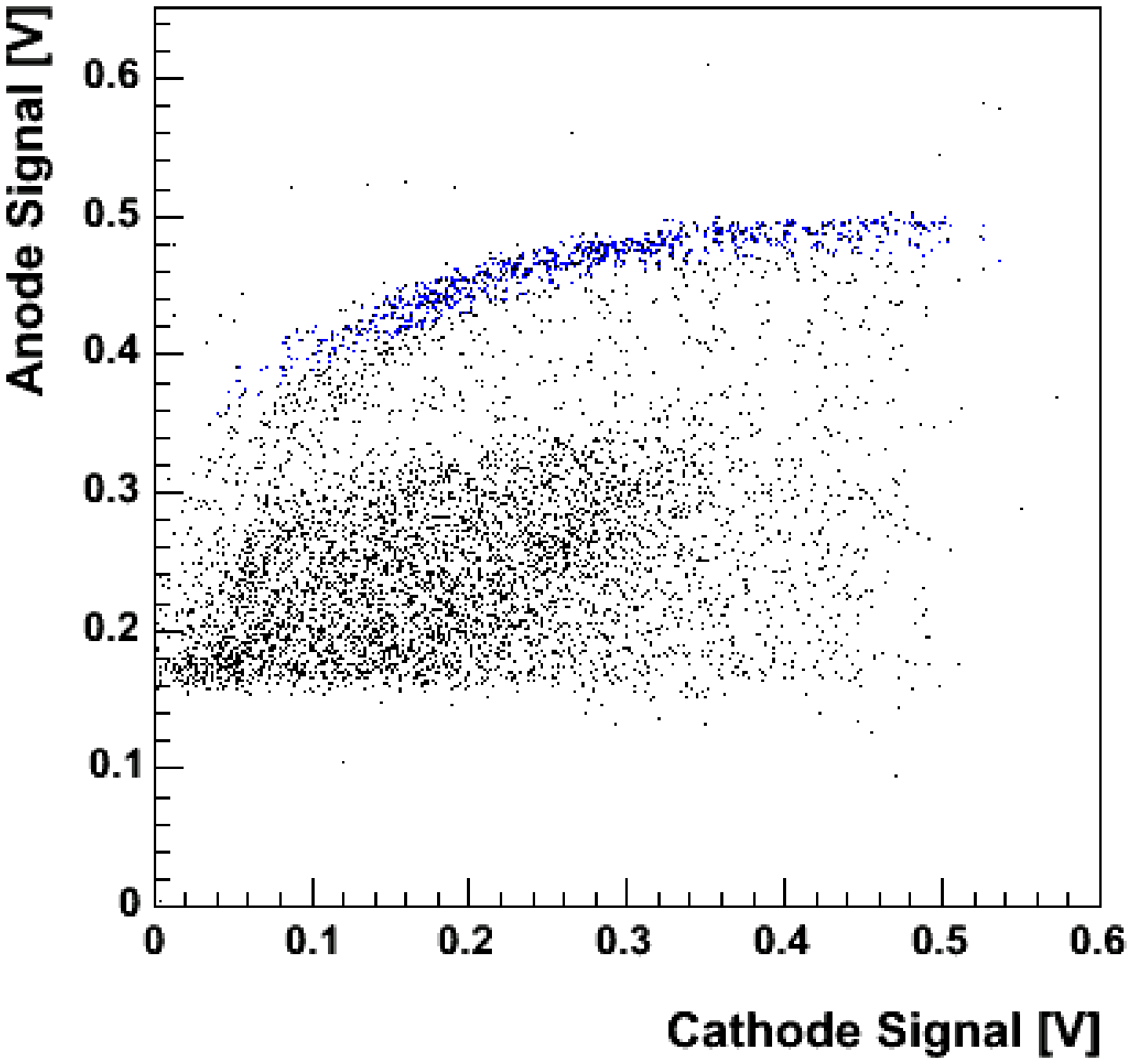}
\vspace*{-0.7cm}
\caption{
The figure shows the anode signal vs. the cathode signal for detector D1b. 
The photopeak events are shown in blue and lie in the upper band. The ratio between the anode signal and the cathode
signal can be used to improve the energy resolution and the number of events
contributing to the photopeak in the energy spectrum.}
\label{DOI}
\end{figure}
The energy resolutions of the detectors were determined using a 662 keV X-ray 
source and cooling the detectors to $-40^\circ$C. The same three pixels for which 
the I-V measurements were performed were also used for measuring the energy spectra. 
A cathode bias voltage of -500 V was applied; at $-40^\circ$C this voltage allowed us to operate
the detectors without incurring breakdown. The pixels were held at ground and AC 
coupled to a fast Amptek 250 amplifier followed by a second-stage amplifier. 
The signals were sampled with a 500 MHz oscilloscope and the digitized pulses were
send to a computer via Ethernet. High-frequency filtering was done offline.
The amplifier noise was 5 keV at -40$^\circ$C and was subtracted in quadrature from
the measured resolutions. In most cases, the amplifier noise was negligible compared
to the inherent detector resolution. A more detailed description of the measurement setup can be found in Krawczynski et al. (2005)
\begin{figure}
\centering
\includegraphics[width=2.5in]{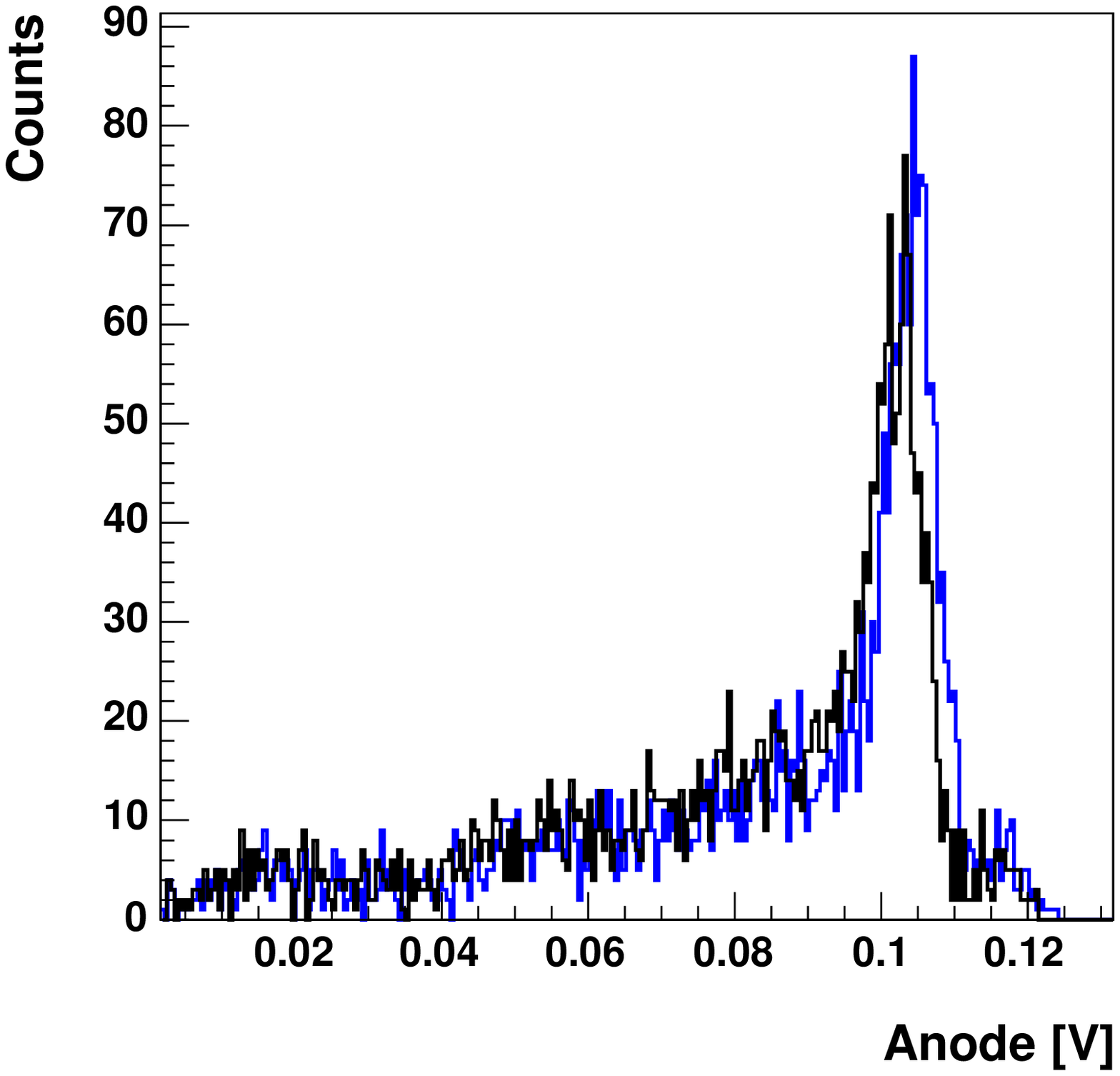}
\vspace*{-0.7cm}
\caption{
Co$^{57}$ (122 keV) spectrum taken with a D1b pixel before (black) and after (blue) correction
for the DOI based on the ratio between anode signal and cathode signal
(anode deposition after polishing, 2 min etching and heat treatment at 90$^\circ$C for 30 min).
After DOI correction the energy resolution is 10 keV (8.3\%).}
\label{Co}
\end{figure}
\begin{figure}
\centering
\includegraphics[width=2.5in]{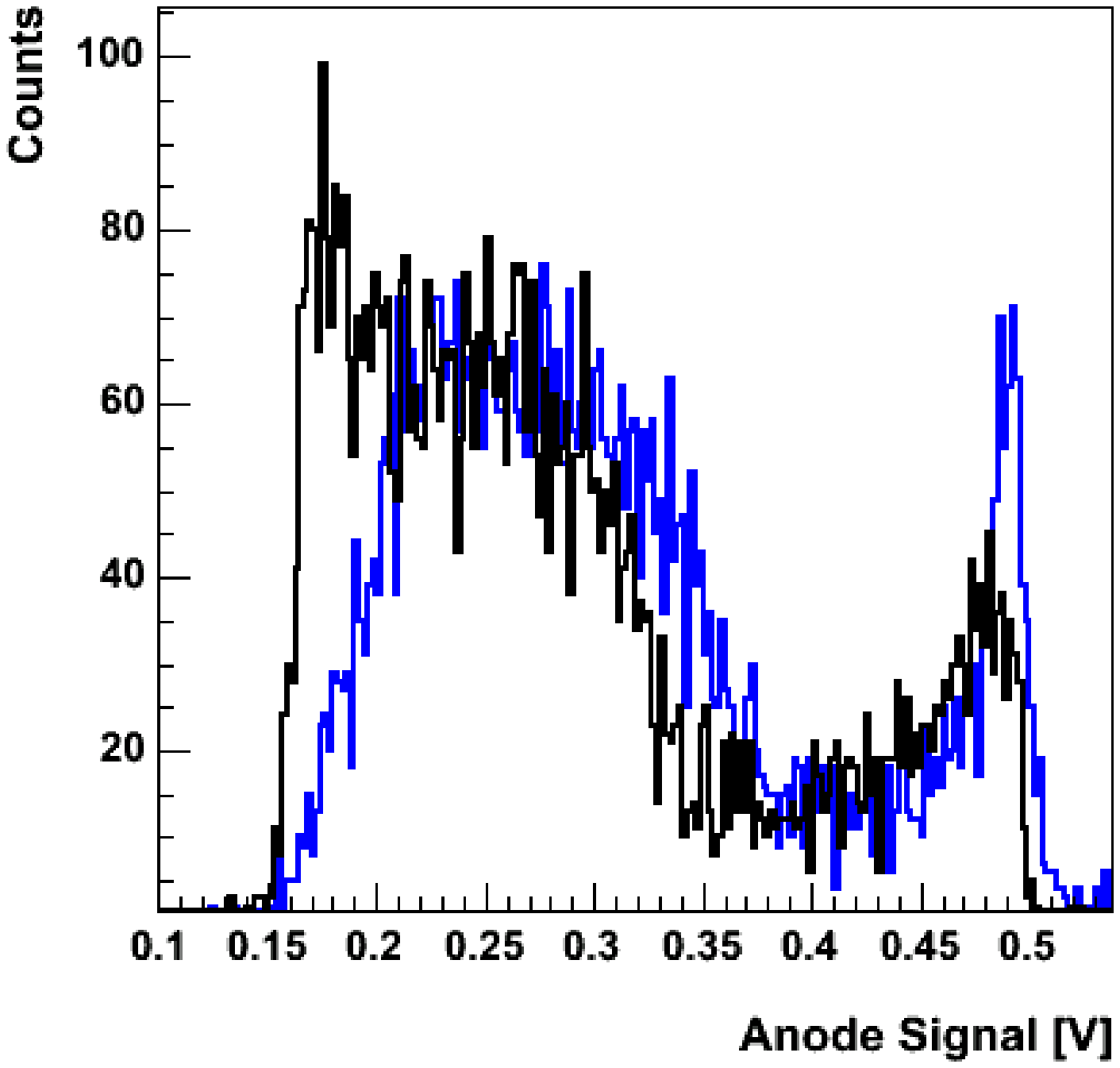}
\vspace*{-0.7cm}
\caption{
Cs$^{137}$ (662 keV) spectrum taken with a D1b pixel before (black) and after (blue) correction
for the DOI based on the ratio between anode signal and cathode signal
(anode deposition after polishing, 2 min etching and heat treatment at 90$^\circ$C for 30 min). 
After DOI correction the energy resolution is 23 keV (3.4\%).}
\label{Cor}
\end{figure}

Owing to trapping of electrons and poor hole mobility, interactions occurring near 
the cathode produce larger anode signals than those that occur near the anodes. 
This depth of interaction (DOI) effect can be corrected by comparing the anode and cathode signals for each pulse. 
Figure 6 shows the anode vs. cathode pulse heights. Photopeak events are highlighted in blue. 
The original and corrected 122 keV and 662 keV energy spectra for a detector D1b pixel 
are displayed in Figs. 7 and 8. After correction we obtain 122 keV and 662 keV full 
width half maximum (FWHM) energy resolutions of 8 keV (6.4\%) and 23 keV (3.4\%), 
respectively.
\begin{table}
\caption{Table 1. Energy Resolutions for Different Contact Materials and Surface Preparation}
\label{table_example}
\centering
\begin{tabular}{|c||c||c||c|}
\hline
Detector & Cathode/ & Surface & Energy \\
 & Anode & Treatment$^a$ & Resolution \\
& & & (662 keV)\\
\hline
D1a & Pt/In & Polish & 8.4\%\\
       &       &        &    \\
\hline
D1b & Pt/In & Polish, Etch(2 min), & 3.4\% \\
       &       & Heat (90$^\circ$C-30 min)&   \\
\hline
D2c & Pt/In & Polish & High Leakage\\
       &       &        & Current \\
\hline
D2a & Au/In & Polish & High Leakage\\
       &       &        &  Current \\
\hline
D2b & Pt/In & Polish, Etch(2 min), & 8.3\% \\
       &       &  Heat (140$^\circ$C,15 min)&  \\
\hline
\end{tabular}
\hspace*{1cm}$^a$ The etching and heat treatment were performed only prior
to depositing the In anodes. The cathodes were always deposited on the 
surfaces as polished.
\end{table}

The energy resolution does depend on the surface treatment and the choice of contact materials. 
Table~1 gives the 662 keV resolutions for the different combinations of substrate, surface treatment,
and contact material. Energy resolutions varied widely from detector to detector and from
pixel to pixel. Here only the results for the best pixels are reported. For all the detectors, we tested three central pixels. The energy resolutions differed from pixel to pixel by $\sim$25\%. It is evident that the combination of etching and heat treatment prior to contact deposition 
improves the performance for both detectors. While the detector D2 did not produce a spectrum with 
any photopeak prior to etching and heating, it did show a photopeak afterward, although the 
662 keV energy resolution (55 keV, 8.3\%) was still very modest. For the detector D1, the 662 keV 
energy resolution improved from 56 keV (8.4\%) before heating and etching to 23 keV (3.4\%) 
after etching and heating. 
\begin{figure}
\centering
\includegraphics[width=2.5in]{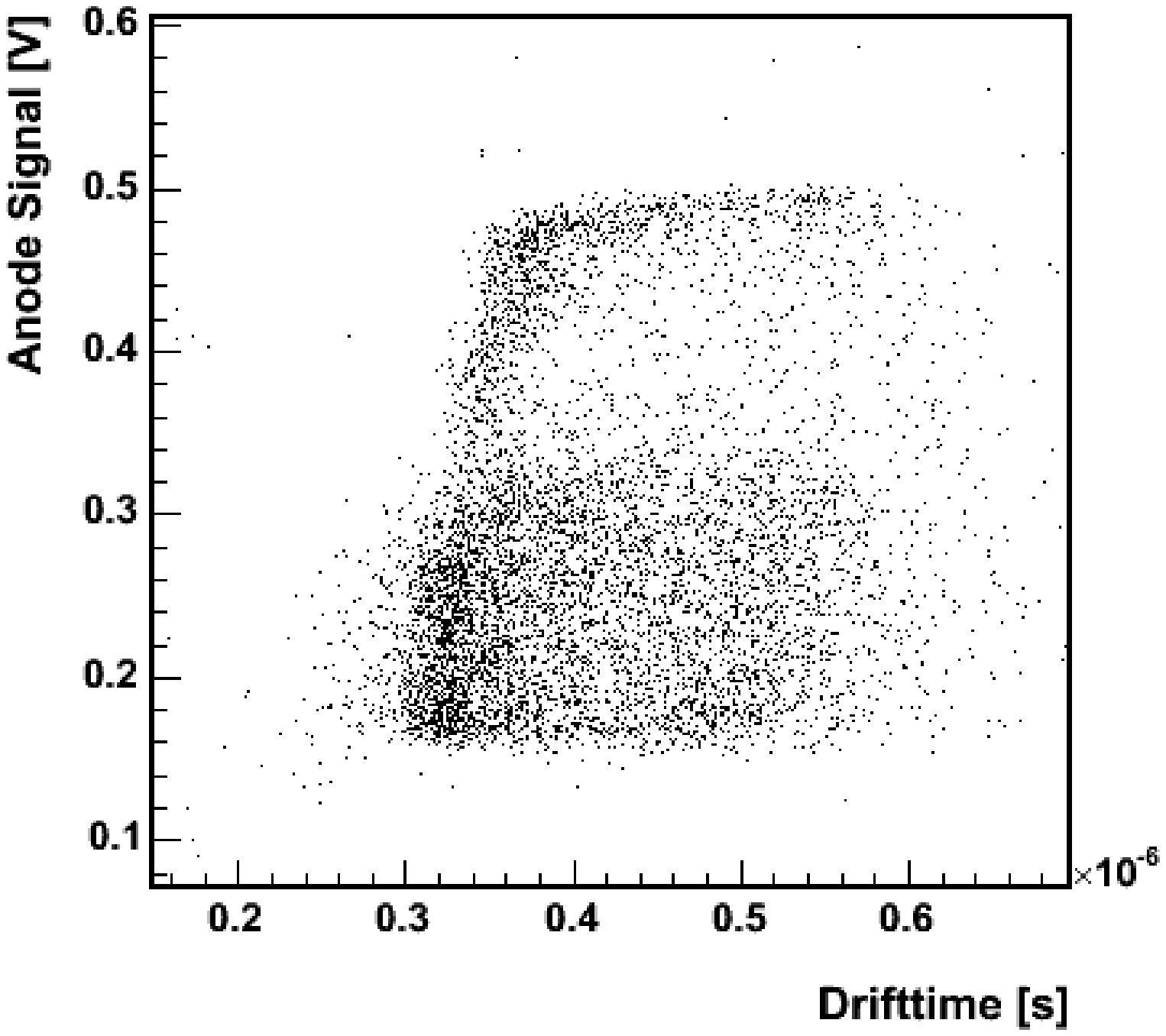}
\caption{
Anode signal versus electron drift times for one pixel of detector D1b
(anode deposition after polishing, 2 min etching and heat treatment at 90$^\circ$C for 30 min).}
\label{Drft}

\end{figure}

Our test-setup allows us to measure the rise time of the pulses and thus to measure
the electron drift times. Assuming  that the detector is fully depleted,
the longest pulse times result from electrons travelling the full 0.5 cm thickness 
of the detector, and the electron mobility $\mu_{e}$ can be computed. 
Figure 9 shows the drift times for detector D1b. 
Longest drift times of $0.58 \times 10^{-6} \rm~sec$ translate into a 
$\mu_{e}$-value of $\sim 860$ cm$^{2}$ V$^{-1}$ s$^{-1}$.
\section{Conclusion}
We have presented here results on the spectroscopic performance of 0.5 cm thick CdTe crystals from the company Acrorad with a variety of electrodes fabricated by Washington Univeristy in St. Louis. At a substrate temperature of -40$^\circ$C, our
best results are 122 keV and 662 keV resolutions of 8 keV (6.4\%) and 23 keV (3.4\%), 
respectively. We have shown that changes in the surface preparation yield widely 
different detector performances. Our detectors showed a breakdown if biased for a 
sufficient time; the time before breakdown was longer for lower bias voltages 
and at colder detector temperatures. We cannot exclude the possibility that a combination 
of low bias voltage and low temperature may suppress the breakdown completely.

Other authors have experimented with between 0.5 and 2.0 mm thick CdTe detectors. 
For example, using a 0.5~mm thick detector with an area of 1.9$\times$1.9 cm$^2$, 
Oonuki et al.\ (2005) report 122 keV and 511 keV energy resolutions of
1.6 keV (1.3\%) and 5.0 keV (1\%). Comparison of these results with 
our results is difficult as Oonuki et al optimized the contact deposition procedure 
with very detailed studies. We thus cannot determine to which degree 
the difference in detector performance results from the different 
detector thicknesses and from the different contacting techniques. 
\appendices
\section*{Acknowledgment}
This work was supported by NASA under contract NNG04WC176.
We thank T. Takahashi for very useful discussions.


%

\end{document}